\documentclass[sigconf]{acmart}
\usepackage{epsfig,endnotes}

\usepackage{mathtools}
\usepackage{amsfonts}
\usepackage{amsmath}
\usepackage{amsthm}
\usepackage{calligra}
\usepackage{bbm}
\usepackage[linesnumbered]{algorithm2e}
\usepackage{xspace}
\usepackage{url}
\usepackage{listings}
\usepackage{enumitem}
\usepackage{times}
\usepackage{epstopdf}
\usepackage{array}
\usepackage{graphicx}
\usepackage{tablefootnote}
\usepackage{comment}
\usepackage{subcaption}
\usepackage[font=small,labelfont=bf]{caption}
\usepackage{multirow}
\usepackage{balance}

\usepackage{booktabs} \usepackage{lipsum} \usepackage{threeparttable} \usepackage{booktabs} \usepackage{makecell} \usepackage{tabularx} \usepackage{amsmath} \usepackage{amssymb} \usepackage{amsthm}

\usepackage{mdwlist}
\usepackage{tgpagella}
\renewcommand{\paragraph}{\vspace{3pt}\noindent\textbf}
\newcolumntype{C}[1]{>{\centering\let\newline\\\arraybackslash\hspace{0pt}}m{#1}}

\newcommand{\name}{{SurFi}\xspace}

\newenvironment{packeditemize}{
\begin{list}{$\bullet$}{
\setlength{\itemsep}{1.5pt}
\setlength{\labelwidth}{8pt}
\setlength{\leftmargin}{10pt}
\setlength{\labelsep}{3pt}
\setlength{\listparindent}{\parindent}
\setlength{\parsep}{1.5pt}
\setlength{\parskip}{1.5pt}
\setlength{\topsep}{1.5pt}}}{\end{list}}
 
\def\BibTeX{{\rm B\kern-.05em{\sc i\kern-.025em b}\kern-.08emT\kern-.1667em\lower.7ex\hbox{E}\kern-.125emX}}
 
\renewcommand\footnotetextcopyrightpermission[1]{}
\begin{document}

\title[\name: Detecting video looping using CSI]{\name: Detecting Surveillance Camera Looping Attacks\\with Wi-Fi Channel State Information}
\subtitle{(Extended Version)\footnote{}\titlenote{This is an extended version of the conference paper in the proceedings of ACM WiSec 2019 with the same title. This is the author's version of the work. It is posted here for your personal use. Not for redistribution. }}

 \author{Nitya Lakshmanan}
 \affiliation{\institution{National University of Singapore}
}
 \email{nityalak@comp.nus.edu.sg}
 
 \author{Inkyu Bang$\dagger$}

 \affiliation{\institution{Agency for Defense Development}
}
\email{ikbang@add.re.kr}
\thanks{$\dagger$ Research done while working at National University of Singapore.}

\author{Min Suk Kang}
\affiliation{\institution{National University of Singapore}
}
 \email{kangms@comp.nus.edu.sg}
 
 \author{Jun Han}
 \affiliation{\institution{National University of Singapore}
}
 \email{junhan@comp.nus.edu.sg}
 
 \author{Jong Taek Lee}
 \affiliation{\institution{Electronics and Telecommunications Research Institute}
}
\email{jongtaeklee@etri.re.kr}

\renewcommand{\shortauthors}{N. Lakshmanan et al.}

\begin{abstract}
The proliferation of surveillance cameras has greatly improved the physical security of many security-critical properties including buildings, stores, and homes. However, recent \textit{surveillance camera looping attacks} demonstrate new security threats\,---\,adversaries can replay a seemingly benign video feed of a place of interest while trespassing or stealing valuables without getting caught. Unfortunately, such attacks are extremely difficult to detect in real-time due to cost and implementation constraints. In this paper, we propose \name to detect these attacks in real-time by utilizing commonly available Wi-Fi signals. In particular, we leverage that channel state information (CSI) from Wi-Fi signals also {\em perceives} human activities in the place of interest in addition to surveillance cameras. \name processes and correlates the live video feeds and the Wi-Fi CSI signals to detect any mismatches that would identify the presence of the surveillance camera looping attacks. \name does not require the deployment of additional infrastructure because Wi-Fi transceivers are easily found in the urban indoor environment. We design and implement the \name system and evaluate its effectiveness in detecting surveillance camera looping attacks. 
Our evaluation demonstrates that \name effectively identifies attacks with up to an attack detection accuracy of 98.8\% and 0.1\% false positive rate.

\end{abstract}
 
\maketitle

\section{Introduction}

Surveillance cameras are now everywhere. Big cities around the world heavily rely on video surveillance to protect themselves from various threats. 
Naturally, video surveillance systems become attractive targets of attacks.
Recent attacks demonstrate surveillance camera looping techniques by launching a software attack on the camera or tapping their Ethernet cables~\cite{BLACKHAT2013_DetectCameraLooping, DEFCON2015_DetectCameraLooping}.
Such new avenue of attacks can potentially enable unauthorized access to a security-sensitive area or unauthorized activities (e.g., stealing valuables or breaking properties) by replaying seemingly legitimate video feeds. 

Unfortunately, mitigation against such attacks is surprisingly challenging. First, many existing legacy surveillance cameras have no proper end-to-end integrity protection and their hardware replacement/upgrade would incur prohibitive costs.
Second, detecting replayed video feeds purely based on video signal analysis is impractical because the surveillance cameras would frequently capture nearly identical but authentic video feeds (e.g., an empty corridor or jewelry shop), which would incur too many false positives. 
Third, the deployment of additional infrastructure for detection (e.g., LED lights blinking in a predefined pattern) will require additional efforts such as secret sharing which potentially introduces a new attack surface.

\begin{figure}[t]
	\centering
	\includegraphics[width=0.49\textwidth]{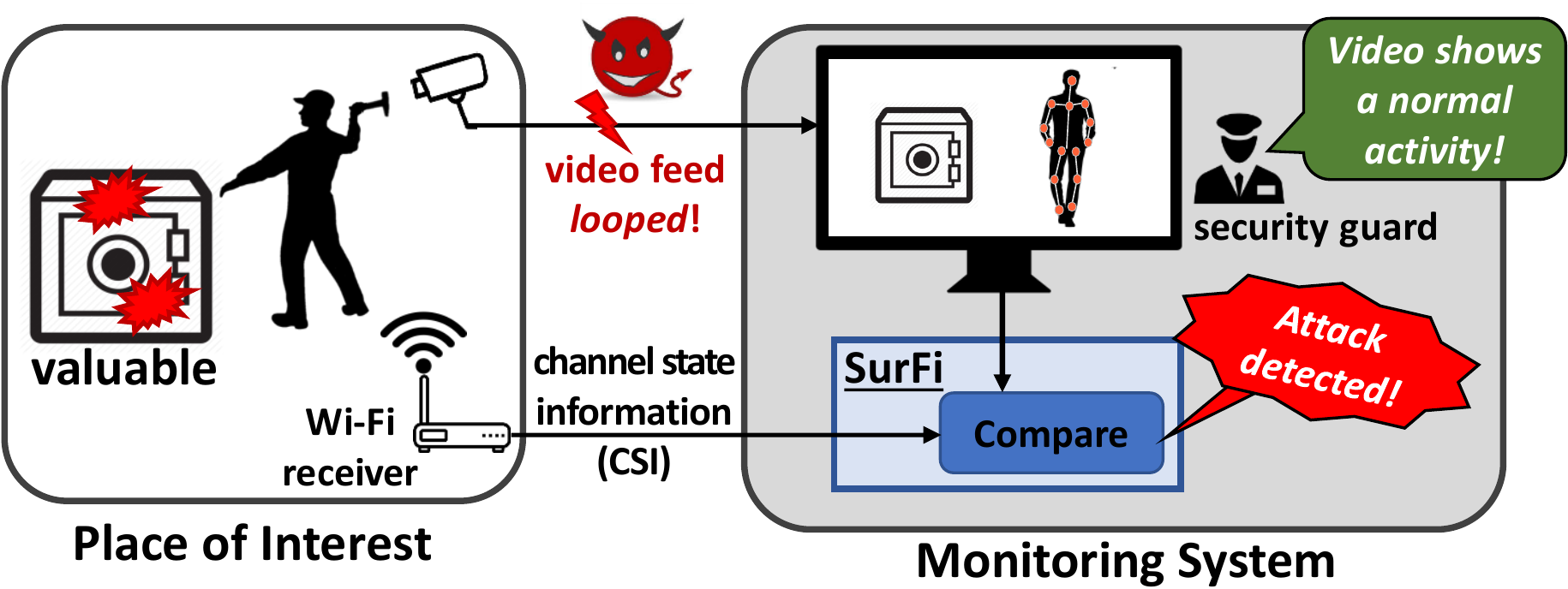} 
\caption{\name detects the surveillance camera looping attack by comparing live video feed with Wi-Fi channel state information (CSI) signal.}
\label{fig:overview}
\end{figure}

In this paper, we try to answer the following question: {\em Can we utilize any auxiliary information from devices that are present along with the surveillance cameras in the place of interest?} 
To answer this question, we investigate whether we can leverage universally deployed Wi-Fi transceivers to effectively detect replayed video feeds. This is based on the observation that co-located devices, though different in sensing modalities, may perceive the same events. 
In this work, we propose \name\footnote{\name stands for `Surveillance with Wi-Fi.'}, which detects in real-time the video feed looping attacks by comparing the video feeds with channel state information (CSI) signals, that can be easily captured via commercial Wi-Fi transceivers. Figure~\ref{fig:overview} exemplifies a scenario where an adversary first loops a seemingly legitimate video feed of an untampered vault, while he is actually breaking into the vault. \name successfully detects this attack in real-time by comparing the looped video feed with the Wi-Fi CSI signal.

Designing \name comes with difficult challenges due to signal comparison across the two different sensing modalities.  First, a direct signal comparison is impossible because signals from heterogeneous sensing modalities (i.e., video and CSI signals) are inherently \textit{semantically different}. In particular, state-of-the-art video processing (such as OpenPose~\cite{2018openpose}) results in displacement of main body points (including head, elbow, waist, and knees) while CSI signals are time-series samples of received wireless symbols across different frequency bands. 
Second, we cannot simply rely on machine learning classification on CSI signals, unlike many research utilizing CSI signals for activity recognition~\cite{2015WiGest,IPSN2016WiWho,wang2015understanding,jiang2018towards}. This is because we \emph{cannot} expect to collect training data for all possible authorized and unauthorized activities of arbitrary visitors of a place of interest 
(e.g., a jewelry shop or a bank ATM office). 

To address these main challenges, we propose to extract common \textit{attributes} from the time-frequency analysis of the signals of differing sensor types. 
Particularly, we observe that the two signals capture the \textit{common timing information} of occurrence of activities and the corresponding \textit{main frequency component}. 
Furthermore, we evaluate our proposal with real-world implementation and demonstrate that \name can detect potential attacks (i.e., non-matching video and CSI pair) with up to an attack detection accuracy (or true positive rate) of 98.8\% and 0.1\% false positive rate. 

 \section{Threat model}
\label{sec:SystemThreatModel}
We first define a {\em place of interest} as an indoor, open space under video surveillance. Some activities in the place of interest are defined unauthorized (e.g., breaking/opening jewelry boxes by unauthorized personnel, breaking/moving ATM machines) and the video surveillance system aims to identify such unauthorized activities. 
We also assume that the surveillance camera's field-of-view covers the entire place-of-interest. This is a reasonable assumption because one installs the video surveillance systems to protect valuable assets, hence purposely minimizing blind spots (i.e., area not in the camera's field-of-view). 

In this paper, we consider the {\em surveillance camera looping attack}, where an adversary is capable of replaying a seemingly legitimate video feed (i.e., containing authorized activities only) to trick the targeted surveillance monitoring system. The ultimate goal of this attack is to evade the detection of the adversary's unauthorized activities by the authorities (e.g., security guards and personnel) as shown in Figure~\ref{fig:overview}.

Recently, security researchers have demonstrated the feasibility of such attacks. In BlackHat 2013~\cite{BLACKHAT2013_DetectCameraLooping}, researchers demonstrated how a vulnerability in the web server interface of a surveillance camera can be leveraged to replace the live video feed with a legitimate-looking image such as an empty store or a lift compartment. In DefCon 2015~\cite{DEFCON2015_DetectCameraLooping}, another group of researchers demonstrated live video feed looping by hijacking an Ethernet cable connection between a camera and a surveillance system without breaking the physical connection. 

Unlike the surveillance camera looping attack, there have been no known CSI measurement looping attacks to date. Hence, we assume that the adversaries cannot loop CSI measurements. \section{System Design}
\label{sec:SystemDesign}

\begin{figure*}[t]
	\centering

	\includegraphics[width=0.95\textwidth]{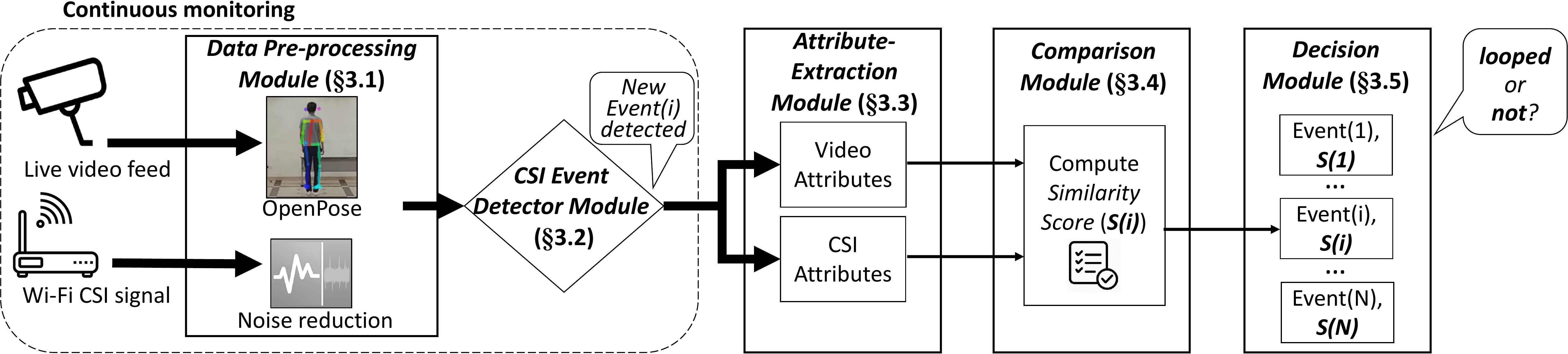} 
\caption{Figure depicts a flowchart illustrating the five modules of \name. The \textit{Attribute Extraction Module} extracts common attributes from different sensing modalities (i.e., video and CSI). The computed scores are then compared in the \textit{Comparison Module}, which in turn is input to the \textit{Decision Module} to output the final attack detection decision.}
	\label{fig:sysDesign}
\end{figure*}
We now describe how \name utilizes the Wi-Fi signal to verify whether the surveillance camera system is under attack. 
Figure~\ref{fig:sysDesign} depicts the flowchart diagram of \name's design consisting of four steps. 
\name continuously receives the real-time video feeds from the surveillance cameras as well as the CSI data from a Wi-Fi receiver.   
The two sets of signals are input to the \textit{Data Pre-processing} module. This module extracts the displacement of body keypoints from the video feeds and removes noise from the CSI signals (\S\ref{subsec:Preprocess}). 
Subsequently, \name monitors the denoised CSI signals to detect the start of an event (\S\ref{subsec:eventdetector}). On detecting the start of an event, the \textit{Attribute Extraction} module extracts three attributes (i.e., start time, end time, and prominent frequency of an event) from both the body keypoints and the denoised CSI signal (\S\ref{subsec:FeatureExtraction}).
Following this, the \textit{Comparison} module compares the attributes and outputs the similarity score (\S\ref{subsec:CorrelationModule}).
Finally, the \textit{Decision} module takes the similarity scores of multiple events and outputs the final decision of whether the video is looped (\S\ref{subsec:decision}).
\begin{figure}[t]
	\centering
	\includegraphics[width=0.49\textwidth]{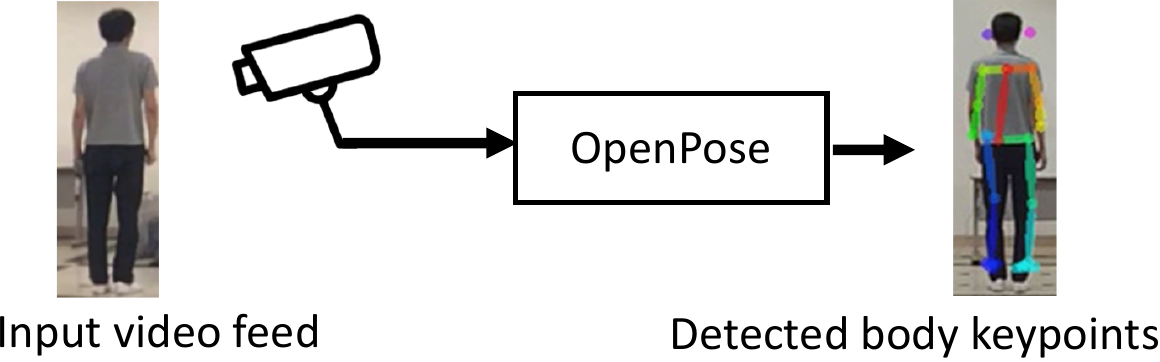} \caption{OpenPose extracts 25 body keypoints (including eyes, hand, feet, joints) and outputs their X-Y coordinate in real time.}
\label{fig:open-pose-example}
\end{figure}

\subsection{Data Pre-processing Module} 
\label{subsec:Preprocess}
In this step, we pre-process the two raw signals obtained from a Wi-Fi receiver and a surveillance camera.The raw CSI signal is composed of multiple time-series CSI values in each subcarrier (e.g., 30 subcarrier per antenna pair in IEEE 802.11n).
We can reliably collect these signals by using various open source tools~\cite{CSItool,MobiSys18_Nexmon}.
We denoise these raw CSI values in order to reliably extract the attributes~\cite{IPSN2016WiWho}.
We reduce noise from a wide frequency range of the received CSI values because we do {\em not} have a pre-specified set of authorized and unauthorized human activities that may occur in a place of interest; i.e., an open space with public access such as ATMs and offices. Hence, we utilize Discrete Wavelet Transform (DWT) filter, which reduces the noise from all frequency bands available in the raw CSI data (e.g., 1--500 Hz in our experiments)~\cite{2015WiGest}.

For the live video feeds, we use OpenPose~\cite{2018openpose}, a state-of-the-art real-time video processing tool, to detect the body keypoints in the video feeds. OpenPose returns the X-Y coordinates of the 25 body keypoints per video frame. Figure~\ref{fig:open-pose-example} shows an example of the body keypoints detected by OpenPose.

\subsection{CSI Event Detector Module}
\label{subsec:eventdetector}
The event detector module receives the denoised CSI values and the X-Y coordinates of the 25 body keypoints, and checks if a start of a new event is detected. This module monitors the CSI values {\em only} because the video feeds can be manipulated by the looping attacks. The start of an event is detected using a metric referred to as \emph{motion energy}~\cite{PhD2008MLHumanActivity}, which is the energy contained in the frequency bands of the CSI signals. When the start of an event is detected, \name triggers the next attribute-extraction module and provide it the denoised CSI values and video keypoints until the end of the event is detected.

\subsection{Attribute-Extraction Module} 
\label{subsec:FeatureExtraction}
The two goals of the attribute-extraction module are (1) to select a set of attributes that enable {\em reliable} comparison of the two signals, and (2) to compute the attributes in {\em real-time}.

\paragraph{Requirements for {\em reliable} comparison.}
The first requirement is that the attributes need to be captured consistently from {\em both} the video and CSI signals. 
Fine-grained information retrievable from video signals (e.g., precise height or gait of a person) may not be adequately extracted from CSI~\cite{IPSN2016WiWho}, and, thus, not suitable for our purpose. 
The second requirement is that the attributes should capture the distinctive characteristics (e.g., the frequency of repeated actions) of {\em activities} with acceptable accuracy. 
One may argue that the accurate identification of individuals will increase the detection of looped video feed (e.g., the live video feed shows Alice whereas the CSI analysis identifies Bob).
Although it may be useful, accurate identification of human beings in a large population (more than 8-10 people) based on the CSI signals remains an open problem~\cite{IPSN2016WiWho, 2016WIFiID}.

\begin{figure}[t]
	\centering
	\includegraphics[width=0.49\textwidth]{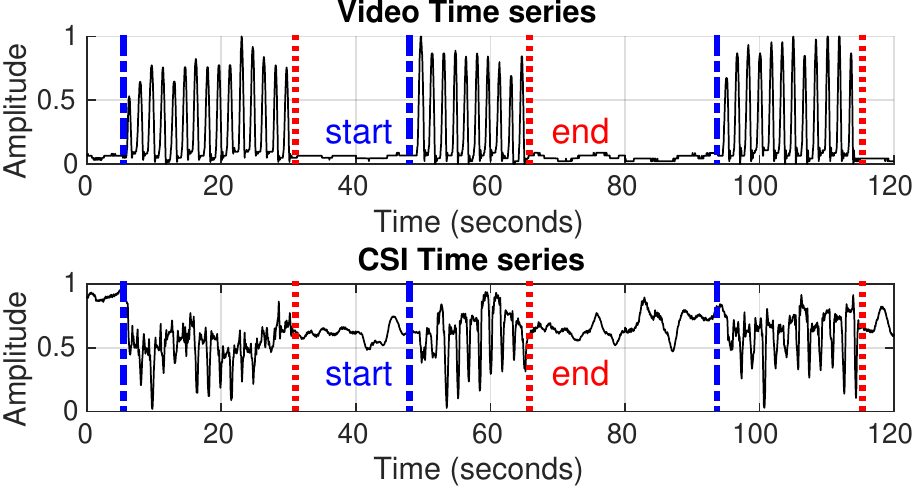} \caption{Time series of the video body keypoint and the selected CSI subcarrier when a person moves his arm up and down. The selected video keypoint is the left wrist and the selected CSI subcarrier is the one with the largest energy component. }
\label{fig:timeseries}
\end{figure}

After extensive experiments under various environments, activities, and human participants, we have found the three attributes that satisfy the above requirements. The first two are time-domain components and the last is a frequency-domain component. 
The time-domain attributes are the \textit{start and end time} of an event whereas the frequency-domain attribute is the \textit{prominent frequency}. When extracting these attributes from the pre-processed video and CSI signals, we choose a single time-series component from each signal, as shown in Figure~\ref{fig:timeseries}. That is, we choose one keypoint with the maximum signal magnitude in the frequency domain from the processed video data, which represents the largest displacement of a single body component during the measurement period. 
Also, we choose one subcarrier signal with the largest energy component from the denoised CSI data.
This simplification reduces the attribute extraction and comparison to a single-dimension problem for real-time analysis while effectively capturing major activities performed in the place of interest. 

\paragraph{Start and end time.} 
We analyze the video and time-domain CSI signals to investigate whether the start and the end time of an activity or an event can be reliably captured. Figure~\ref{fig:timeseries} depicts an activity where a person moves his left arm up and down for a random time duration, repeating it three times. 

Note that similar video signals can be obtained from the other body parts (e.g., the left elbow) but our algorithm picks the left wrist as it contains the largest magnitude. 
Both signals exhibit a similar start and end time, and this trend is consistently observed over multiple trials, people, and events. This allows us to conclude that they can be the reliable attributes for comparison between the video and CSI signals.  

To extract the start and the end time of an event, we use a metric referred as \emph{motion energy}~\cite{PhD2008MLHumanActivity} which captures the energy in the different frequency bands of the CSI signal. The motion energy ($E$) can be calculated as
 \begin{align}
E  = \sum_{i=1}^{L} FFT_{half}(i)^2,
 \label{eq:motion-energy}
 \end{align}
where the $FFT_{half}(i)$ is the FFT coefficient magnitude calculated over a time window, $L$. Note that we only consider the first half of the FFT coefficients since other half is redundant and we ignore the DC component. We divide the total event time window into $L$ of 0.1 seconds for both video and CSI signals since this gives sufficient time granularity for comparison of attributes. 

We detect the start time of an event when its motion energy $E$ increases and crosses a threshold.
For the video feeds, we monitor the instantaneous $E$ values and determine the start of an event when $E$ becomes ten times larger than its moving average value. For CSI signals, which is still much noisier than the video signals, we determine the start of an event when the variance of $E$ becomes larger than its moving average. The end times can be calculated similarly. 
We denote the start and the end times measured for an event from video and CSI signals as follows: $\tau_{\tt start}^{\tt V}$, $\tau_{\tt end}^{\tt V}$, $\tau_{\tt start}^{\tt C}$, and $\tau_{\tt end}^{\tt C}$.

\begin{figure}[t]
	\centering
	\includegraphics[width=0.49\textwidth]{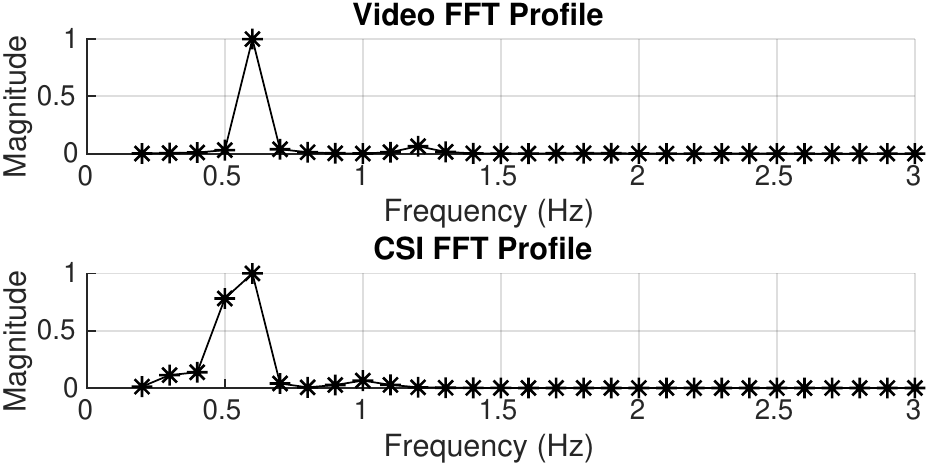} \caption{FFT profile of the left wrist body keypoint and the selected CSI subcarrier when a person moves his arm up and down. The prominent frequency of both the signals is 0.6 Hz, which is the frequency at which the event is performed.}
\label{fig:freq}
\end{figure}

\paragraph{Prominent frequency.} 
The third attribute we extract is the prominent frequency component of an event. 
We observe that it is possible to extract a single frequency component that exhibits the maximum signal magnitude in the frequency domain.
Figure~\ref{fig:freq} shows the FFT plot of the selected video and CSI signals of the aforementioned activity discussed in Figure~\ref{fig:timeseries}. We observe that both the video and CSI signals exhibit approximately 0.6 Hz main frequency component. 
Hence, we conclude that a repeated event is captured as the prominent frequency component in the frequency domain. 

To calculate the prominent frequency of an event, we first apply FFT on the video signal and consider the frequency with the maximum magnitude as the prominent frequency, $f^{\tt V}$. Then, we apply a bandpass filter on the CSI signal with the frequency we obtain from the video signal $f^{\tt V}$ to remove unrelated frequency components (e.g., less than 0.3 Hz which mainly captures slower movements such as posture changes~\cite{IPSN2016WiWho}) and extract the maximum magnitude frequency, $f^{\tt C}$.

\subsection{Comparison Module}
\label{subsec:CorrelationModule}
Given the two sets of attributes \{$\tau_{\tt start}^{\tt V}$,  $\tau_{\tt end}^{\tt V}$ , $f^{\tt V}$\} and \{$\tau_{\tt start}^{\tt C}$,  $\tau_{\tt end}^{\tt C}$, $f^{\tt C}$\} from the video and CSI signals, we compute the similarity score $S(i)$ for a single observed event $Event(i)$. 

We first determine the {\em per-attribute} similarity scores between the two signals as follows,
\begin{equation}
  AS_{j} =
    \begin{cases}
      1 & \text{if $\triangle_{j} \leq T_{j}$}\\
      0 & \text{otherwise,}
    \end{cases}
    ,~~j \in \{1, 2, 3\},
\end{equation}
where $T_{j}$ is the per-attribute threshold and $\triangle_{j}$ is the difference between the three attributes of the two signals as follows:
$\triangle_{1}$ = $|\tau_{\tt start}^{\tt V} - \tau_{\tt start}^{\tt C}|$,
$\triangle_{2}$ = $|\tau_{\tt end}^{\tt V} - \tau_{\tt end}^{\tt C}|$, and
$\triangle_{3}$ = $|f^{\tt V} - f^{\tt C}|$ .
The per-attribute thresholds are chosen empirically to obtain an acceptable detection accuracy with a low false positive rate per attribute. To achieve this, we compute the accuracy with which each attribute detects a legitimate video as legitimate, and a looped video as looped. A low threshold value may result in misidentifying the legitimate video as looped, whereas a high threshold value will cause the looped video to be incorrectly detected as legitimate, thereby decreasing the accuracy. Thus, the threshold at which the accuracy is maximized is selected as the per-attribute threshold. Based on our experiments, we observe the start and end time attribute thresholds of 2.5 seconds and 2 seconds, respectively, and a frequency attribute threshold of 0.25 Hz to be optimal.
Subsequently, we compute the {\em per-event similarity score}, which ranges from 0 to 3 as follows,
 \begin{equation}
S(i) = \sum_{j=1}^{3} AS_{j}.
 \end{equation}

\subsection{Decision Module}
\label{subsec:decision}
The decision module takes as input the similarity scores of one or more individually observed events and outputs a final decision of whether the video feed is looped. When multiple events are considered, detection accuracy (i.e., how effectively \name detects when the video feed is looped) improves and false positive rate (i.e., how frequently \name misidentifies a legitimate case as an attack) decreases. This is because as more similarity scores from independently observed events are considered, \name achieves higher confidence level for the final decision. We first take the average of the similarity score of multiple events and compare it against a decision threshold. The choice of this threshold affects the detection accuracy and the false positive rate. We decide to set a low false positive rate for \name and then evaluate the detection accuracy.
 \section{Evaluation}
\label{sec:Evaluation}
We evaluate the feasibility of \name by conducting preliminary experiments in a controlled environment. We describe our experiment setup and present the performance evaluation.

\subsection{Experiment Setup}
\label{subsec:setup}
\begin{figure}[t]
	\centering
	\includegraphics[width=0.49\textwidth]{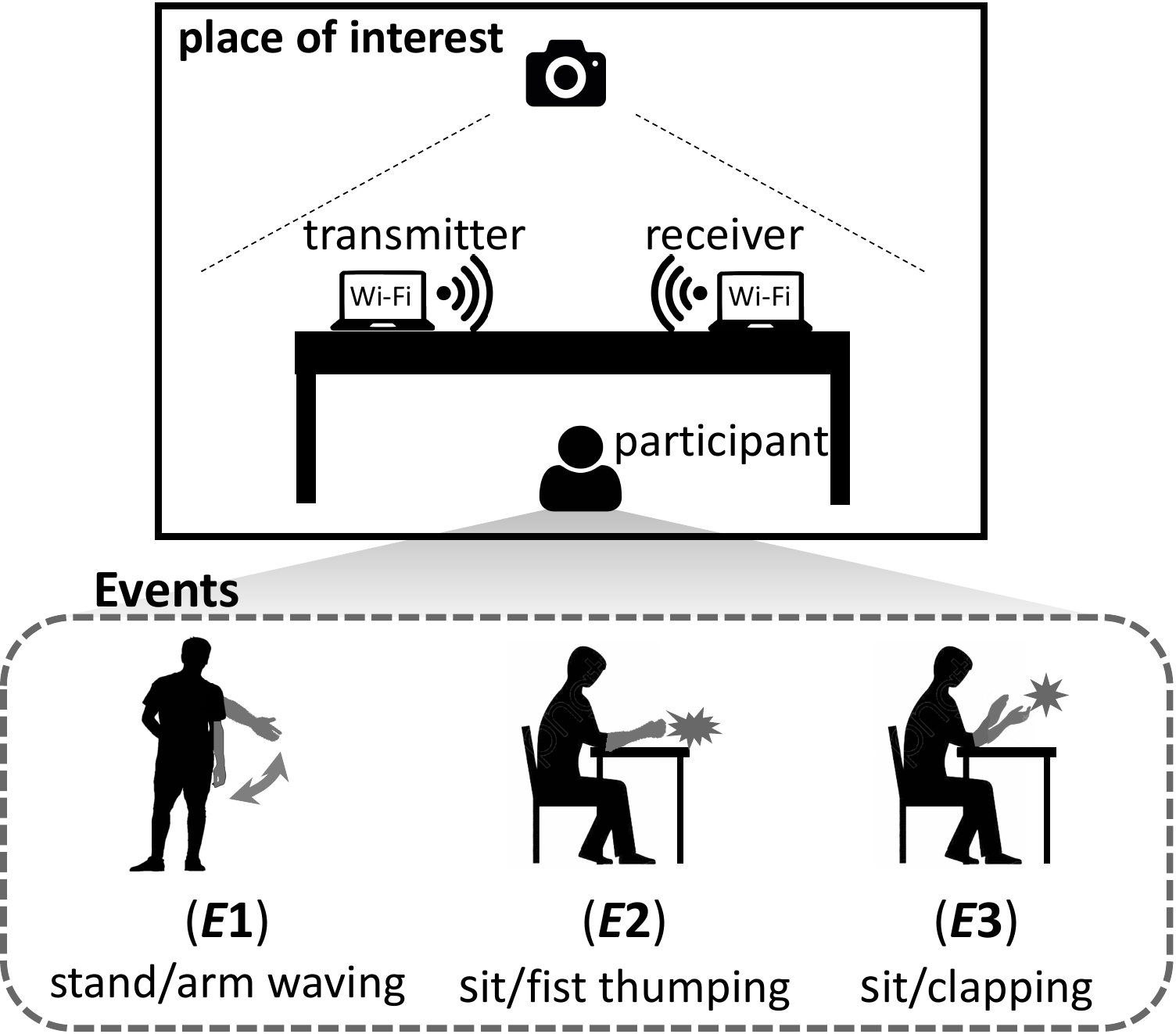} \caption{Experiment setup depicting the location of the camera recording the events performed, and the two Wi-Fi transceivers used for collecting CSI, and the three events we test.}
\label{fig:setup}
\end{figure}

Figure~\ref{fig:setup} shows the setup of our place of interest. We set a controlled experiment in a small indoor office room (4.9-meter wide $\times$ 2.6-meter long) . 
We set up a transmitter-receiver pair using two Thinkpad W500 laptops equipped with Intel NIC 5300, which are spaced 1.6-meter apart on the table.
We use Linux 802.11n CSI Tool~\cite{CSItool} on Ubuntu 14.04 to extract CSI values from Wi-Fi packets at 5~GHz frequency range. The transmitter sends one ping packet every 1~millisecond and the receiver collects the CSI values. Overall, we obtain the CSI values of 90 subcarriers as the two laptops are equipped with one transmitter and three receiver antennas.

For video feed collection, we record video clips (at 13-Megapixel resolution at the 30 frames/second sampling rate) using a camera on a mobile phone installed in the middle of the wall, with the field-of-view that faces a participant.

We ask the participants to perform the following three events:
\begin{itemize}
    \item [({\bf $E1$})] standing and moving the left arm up and down repeatedly at 0.6 Hz frequency;
    \item [({\bf $E2$})] sitting and thumping right fist on the table repeatedly at 1~Hz frequency; and 
    \item [({\bf $E3$})] sitting and clapping his/her hands repeatedly at 1.6 Hz frequency.
\end{itemize}

In our experiments, we test the three events with four participants and each event is tested 21 trials per participant. We instruct the participants to perform the selected activity at a certain frequency. 
To represent more realistic event occurrences, we randomize the starting and ending time of each trial (chosen uniformly at random between 5 to 15 seconds range, and 25 to 35 seconds range, respectively) such that each trial length ranges from 10 to 30 seconds. 

\subsection{Attribute-based Similarity Comparison}
\label{subsec:Feasibility}
\begin{figure}[t]
	\centering
	\includegraphics[width=0.45\textwidth]{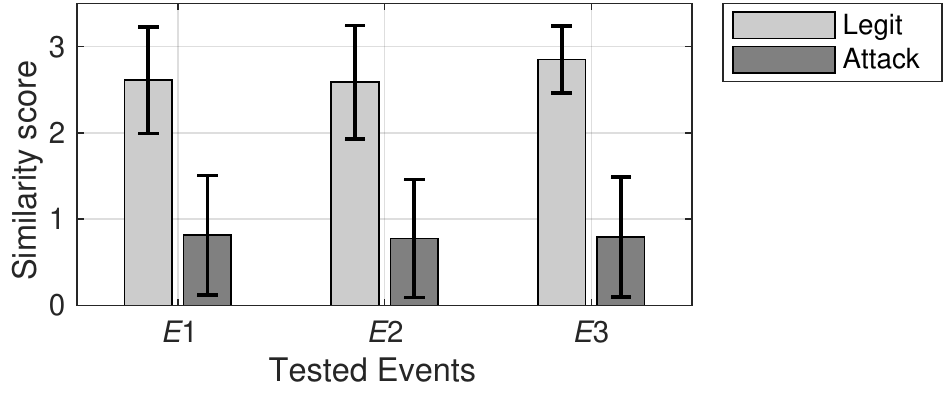} \caption{The distribution of similarity scores of the three events for the legitimate and attack cases. 
	In the  \emph{legitimate} cases, we compare a \emph{live} video feed and the CSI signal of the same event.
	In the \emph{attack} cases, we instead use a \emph{looped} video feed with a different event.
	The bars represent the average scores whereas the error bars denote the standard deviations. }
\label{fig:averagescore}
	
\end{figure}

We now evaluate how \name utilizes the three attributes to compare similarities between the video and CSI signals, ultimately to detect for any potential attacks. Figure \ref{fig:averagescore} depicts the distribution of the per-event similarity scores $S(i)$ of an observed $Event(i)$ for the legitimate and attack cases across the three event types. 
For the legitimate cases, we compute $S(i)$ by comparing the attributes extracted from the live (thus authentic) video feed and the CSI signal. Subsequently, we treat the remaining cases as attack cases. Hence, for the attack cases, we compute $S(i)$ by comparing the attributes extracted from the looped video feeds, which contain different event types, and the CSI signal; e.g., $E1$ is performed in a looped video feed while $E2$ is performed in the actual experiment. 
Overall, we have 84 trials per event type (e.g., $E1$, $E2$, $E3$) representing the legitimate cases and 14,112 trials per event type representing the attack cases. 
The legitimate cases show much higher average per-event similarity scores (i.e., 2.6, 2.5, and 2.8) than the attack cases (i.e., 0.81, 0.77 and 0.79). 
The clear difference in the per-event similarity scores of the two cases shows that the selected attributes can effectively differentiate the cases independently of event types and subjects.
Yet, their high variance (e.g., often close to 1) may cause some incorrect attack detection and false alarms.

\subsection{Attack Detection Accuracy}
\label{subsec:MultiEvent}
We now evaluate the overall performance of \name when multiple events are observed within a short time duration (e.g., a few minutes) to increase the attack detection accuracy. 
We define two performance metrics as follows:
\begin{packeditemize}
    \item[(1)] {\em Attack detection accuracy} (or true positive rate): a ratio that the looped video feeds are correctly identified as looped;
    \item[(2)] {\em False positive rate}: a ratio that the legitimate video feeds is incorrectly detected as looped. 
\end{packeditemize}
\begin{figure}[t]
	\centering

	\includegraphics[width=0.45\textwidth]{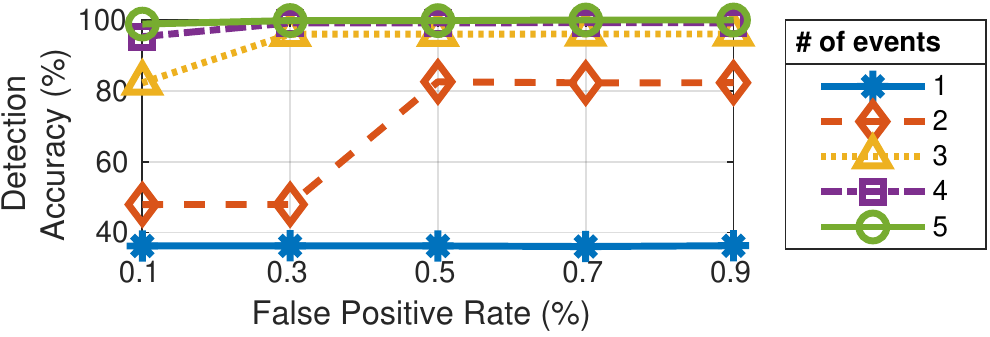}
\caption{The attack detection accuracy improves as \name uses more events for attack detection. }
\label{fig:MultiEventPlot}
\end{figure}

We randomly append events together to emulate multiple event sequence. We then take the average of per-event similarity scores of the sequence. As discussed earlier, the more events we observe and use to detect the attacks, the higher confidence we have for our final decision, thus improving the attack detection accuracy. 
Figure~\ref{fig:MultiEventPlot} shows the attack detection accuracy of the \name's attack detection for a different number of events varying from 1 to 5 and different target false positive rates (0.1\% to 0.9\%). 
We can observe that \name achieves 36\% detection accuracy with only a single event when we target the false positive rate no higher than 0.1\%. It can achieve up to 98.8\% detection accuracy when it uses five events for the same target false positive rate. As the target false positive rate increases, the detection accuracy further increases as depicted in the figure. 

  \begin{figure}[t]
  	\centering
\includegraphics[width=0.45\textwidth]{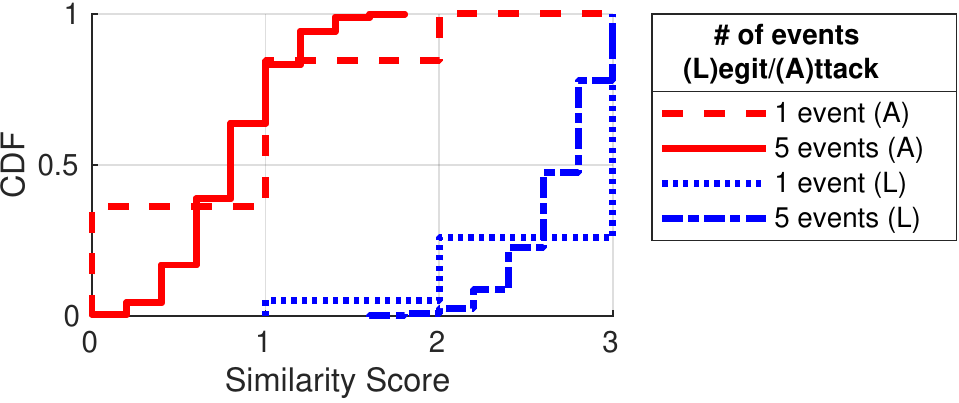} \caption{Cumulative distribution function (CDF) of legitimate and attack cases for different numbers of events used for attack detection.} 
\label{fig:MultiEventCDF}
  \end{figure}

  Furthermore, Figure~\ref{fig:MultiEventCDF} explains why we can achieve higher detection accuracy as we use more events for attack detection. The figure shows the CDF of the similarity scores when the number of events is 1 and 5. It can be observed that for both legitimate and attack cases, the variance of the similarity score distributions decreases as the number of events increases. 
 \section{Discussion}
\label{sec:Discussion}
In this section, we discuss the potential improvements for future work and some of the deployment considerations.

\paragraph{Stronger adversary model.} One of the ways that an attacker could defeat \name is to mimic a subject's activities played in the compromised video. However, the level of sophistication required for such an attack would be significant, as small differences will cause differences in the \textit{attributes}. Furthermore, \name's \textit{Decision Module} requires the attacker to consistently mimic multiple events in series, making circumventing the \name defense even harder. 
We leave the study of more sophisticated attacks for future work.

\begin{figure}[t]
 	\centering
 	\includegraphics[width=0.45\textwidth]{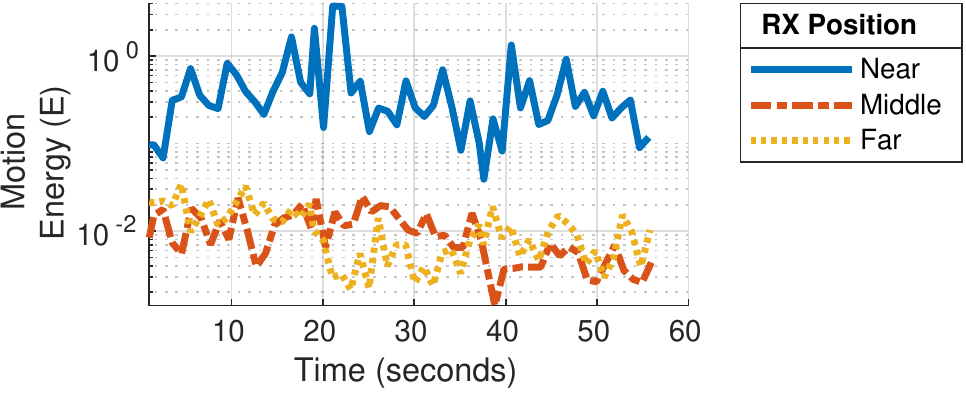} \caption{Behind-the-wall activity effects the CSI measurement only when the receiver (RX) position is near the activity being performed outside.} 

 	\label{fig:Outside}
 \end{figure}

 \paragraph{Behind-the-wall activities.} One remaining concern is that the \name's performance may degrade (i.e., high false positives) because of the activities outside the physical perimeter of our place of interest. 
 However, our simple experiments shows that behind-the-wall activities are not detected as long as they are far away from the receiver (e.g., 3.6-meter).
 We evaluate the performance of \name by conducting experiments at varying positions of the receiver. Figure~\ref{fig:Outside} shows the motion energy calculated from the CSI measurements for three different receiver positions: receiver is at a distance of 5-meter (\textit{far}), 3.6-meter (\textit{middle}), and 0.9-meter (\textit{near}) from the corridor outside the experiment room in Figure~\ref{fig:setup}.

 The plot clearly shows that for the receivers located in \textit{far} and \textit{middle} positions, the behind-the-wall activities are not detected since the corresponding motion energy is close to zero due to signal attenuation of the wall. In contrast, for the receiver located in the \textit{near} position, the varying motion energy often leads to false detection of an event. This is because the receiver close to the corridor (e.g., 0.9-meter) can still measure the CSI and detect the event even if the CSI signal suffers attenuation.

\paragraph{Deployment considerations.} When the \name system is deployed in practice, one may need to perform some {\em calibrations} because some thresholds used in our attack detection (e.g., thresholds for per-attribute similarity scores in Section~\ref{subsec:CorrelationModule} or multiple-event based decision in Section~\ref{subsec:decision}) should be adjusted to the new environment. 

    \section{Related Work}
\label{sec:related}
This section presents related work on wireless sensing as well as cross-modality sensing. 
Refer to Section~\ref{sec:SystemThreatModel} for the recent work on video camera looping attacks. 

\paragraph{Wireless Sensing.} Many researchers demonstrate the feasibility of leveraging wireless signal (including Wi-Fi) for recognizing various human movements including human activity and gesture recognition~\cite{2014WiTrack,2013SeeTW,PhD2008MLHumanActivity,2013WiSee,2015WiGest} as well as person identification using gait patterns~\cite{IPSN2016WiWho,UBICOMP2016GaitAnalysis,2016WIFiID}. All of these research utilize machine learning classification that requires the collection of training data. While \name also utilizes the analysis of CSI signals, we address an inherently more challenging problem as we cannot expect to collect training data. Bagci \emph{et al.}~\cite{AACSA2015_CSI_Tampering} considers the problem of physical tampering of the Wi-Fi enabled cameras (e.g., moving or rotating the surveillance camera), which is different from our problem.

\paragraph{Cross-modality Sensing.} 
Recent body of work studies how to utilize heterogeneous sensing modalities for added benefits. Researchers correlate video camera image with other sensing modalities such as RSSI value or IMU data to provide additional verification in different applications~\cite{2014IdentityLink,idrone,universense}. We are inspired by these techniques but \name addresses an inherently more difficult challenge of performing cross-modality correlation without the need of deploying specific sensors on users or objects, but rather utilizing the sensed information from existing infrastructure.
 \section{Conclusion}
\label{sec:conclusion}
    In this paper, we propose \name, a system that detects \textit{surveillance camera looping attack} in real-time. \name leverages existing Wi-Fi infrastructure (thus requiring no additional hardware or deployment costs) to extract channel state information (CSI) to process and correlate video and CSI signals to detect any mismatches. \name increases its detection confidence as more events are perceived and correlated by the two heterogeneous sensing modalities. Our proof of concept of \name achieves up to an attack detection accuracy of 98.8\% and false positive rate of 0.1\%. As our future work, we plan to conduct more comprehensive experiments with diverse events and environments against a more sophisticated adversary model.

\section{ACKNOWLEDGMENTS}
This research was partially supported by the grants from Singapore Ministry of Education Academic Research Fund Tier-1 (R-252-000-690-114, R-252-000-A26-133). 


\end{document}